\documentclass[aps,superscriptaddress,amsmath,amssymb,twocolumn,floatfix,english]{revtex4}

\usepackage{url}
\usepackage{bm}
\usepackage{graphicx}
\usepackage{amsmath}
\usepackage{amstext}
\usepackage{amssymb}
\usepackage{amsfonts}
\usepackage{amsbsy}
\usepackage{verbatim}
\usepackage{color}
\usepackage[colorlinks=true, urlcolor=blue, linkcolor=blue, citecolor=blue, pdftex]{hyperref}
\usepackage{multirow}
\usepackage{pdfpages}
\usepackage{floatrow}
\usepackage{gensymb}
\usepackage{textcomp}

\newcommand{\dagga}{{\phantom{\dagger}}}

\begin{document}

\title{The Hubbard model on triangular $N$-leg cylinders: chiral and non-chiral spin liquids}

\author{Luca F. Tocchio}
\affiliation{Institute for Condensed Matter Physics and Complex Systems, DISAT, Politecnico di Torino, I-10129 Torino, Italy}
\author{Arianna Montorsi}
\affiliation{Institute for Condensed Matter Physics and Complex Systems, DISAT, Politecnico di Torino, I-10129 Torino, Italy}
\author{Federico Becca}
\affiliation{Dipartimento di Fisica, Universit\`a di Trieste, Strada Costiera 11, I-34151 Trieste, Italy}

\date{\today}

\begin{abstract}
The existence of a gapped chiral spin liquid has been recently suggested in the vicinity of the metal-insulator transition of the Hubbard 
model on the triangular lattice, by intensive density-matrix renormalization group (DMRG) simulations [A. Szasz, J. Motruk, M.P. Zaletel, 
and J.E. Moore, Phys. Rev. X {\bf 10}, 021042 (2020)]. Here, we report the results obtained within the variational Monte Carlo technique 
based upon Jastrow-Slater wave functions, implemented with backflow correlations. As in DMRG calculations, we consider $N$-leg cylinders. 
For $N=4$ and in the presence of a next-nearest neighbor hopping, a chiral spin liquid emerges between the metal and the insulator with 
magnetic quasi-long-range order. Within our approach, the chiral state is gapped and breaks the reflection symmetry. By contrast, for both 
$N=5$ and $N=6$, the chiral spin liquid is not the state with the lowest variational energy: in the former case a nematic spin liquid is 
found in the entire insulating regime, while for the less frustrated case with $N=6$ the results are very similar to the one obtained on 
two-dimensional clusters [L.F. Tocchio, A. Montorsi, and F. Becca, Phys. Rev. B {\bf 102}, 115150 (2020)], with an antiferromagnetic phase 
close to the metal-insulator transition and a nematic spin liquid in the strong-coupling regime.
\end{abstract}

\maketitle

\section{Introduction}\label{sec:intro}

The quest for spin-liquid states has fascinated the condensed matter physics community since the first proposal of the resonating-valence 
bond (RVB) theory by Fazekas and Anderson~\cite{anderson1973,fazekas1974}. This approach has been one of the first attempts to describe a 
Mott insulator without any sort of symmetry breaking, even at zero temperature. In recent years, spin liquids have been reported in an 
increasing number of materials. Examples are given by Herbertsmithite, which is well described by the Heisenberg model on the kagome 
lattice~\cite{norman2016}, organic compounds, like $\kappa$(ET)$_2$Cu$_2$(CN)$_3$ and Me$_3$EtSb[Pd(dmit)$_2$]$_2$~\cite{kanoda2011,powell2011}, 
or transition-metal dichalcogenides, like 1T-TaS$_2$, whose low-temperature behavior could be captured by the Hubbard model on the triangular 
lattice~\cite{law2017}.

An important open question concerns the nature of the insulating phase of the two-dimensional Hubbard model on the triangular lattice at half 
filling. Most of the investigations have been concentrated to its strong-coupling regime, where only spin $S=1/2$ degrees of freedom are left. 
Here, spin liquids can be systematically classified, according to the projective-symmetry group (PSG) theory~\cite{zhou2002,wang2006,lu2015}. 
In particular, one can distinguish between $\mathbb{Z}_2$ and $U(1)$ spin liquids, according to the low-energy symmetry of the emerging gauge
fields~\cite{wen2002}. Starting from the Heisenberg model with nearest-neighbor (NN) super-exchange $J$, spin-liquid phases are expected to 
be stabilized when including either a next-nearest-neighbor (NNN) coupling $J^\prime$ or a four-spin ring-exchange term $K$. The latter one 
can be justified within the fourth-order strong-coupling expansion in $t/U$ and is usually considered for an effective description of density 
fluctuations close to the Mott transition~\cite{grover2010}. As far as the $J-J^\prime$ model is concerned, a gapless $U(1)$ spin liquid has 
been proposed by both variational Monte Carlo (VMC)~\cite{kaneko2014,iqbal2016} and recent density-matrix renormalization group (DMRG) 
calculations~\cite{hu2019}, while older DMRG results suggested the presence of a gapped spin liquid~\cite{zhu2015,hu2015}. In addition, also
ring-exchange terms may stabilize a gapless spin liquid (with a spinon Fermi surface), as proposed by earlier VMC studies~\cite{motrunich2005} 
and confirmed by later DMRG simulations~\cite{block2011,he2018}. Further VMC investigations suggested two other gapless spin-liquid states, 
none of them possessing a spinon Fermi surface~\cite{mishmash2013}. However, more recent tensor-network approaches, implemented from 
Gutzwiller-projected wave functions, do not support the existence of a gapless spin liquid~\cite{aghaei2020}.

Recently, chiral spin liquids attracted much attention, because of their similarities with quantum Hall states~\cite{kalmeyer1987,wen1989}.
Interestingly, chiral states may exist not only when the Hamiltonian explicitly breaks time-reversal symmetry (as in the quantum Hall 
effect)~\cite{bauer2014}, but also as a result of a spontaneous symmetry breaking phenomenon~\cite{he2014}. On the triangular lattice, some 
evidence of this exotic phase has been obtained by adding a scalar chiral interaction to the Heisenberg Hamiltonian~\cite{hu2016,wietek2017},
or even in a fully symmetric Heisenberg model with super-exchange couplings up to the third neighbors~\cite{gong2019}. A PSG classification 
of chiral states is possible, as worked out for the fermionic case for different lattices~\cite{bieri2016}. In particular, two simple 
{\it Ans\"atze} can be constructed~\cite{song2020}: The first one (dubbed CSL1) is a $U(1)$ chiral spin liquid, with complex hoppings 
defined on a $2 \times 1$ unit cell and no pairing; the second one (dubbed CSL2) is a Gutzwiller projected $d+id$ superconductor. 

The situation in the Hubbard model, characterized by a kinetic term $t$ and an on-site Coulomb repulsion $U$, is much less clear. The main 
difficulty comes from the presence of density fluctuations, whose energy scale is related to $U$ that is much larger than the typical energy 
scale of spin fluctuations, i.e., $J=4t^2/U$. Therefore, it is not simple to detect tiny effects related to spin degrees of freedom when 
density fluctuations are present. In addition, numerical methods like exact diagonalization or DMRG suffer from the fact that the local 
Hilbert space is doubled with respect to the case of $S=1/2$. Nevertheless, this effort is necessary in order to capture density fluctuations 
that are inevitably present in real materials. The possibility that a spin-liquid phase may exist not only in the strong-coupling regime 
$U \gg t$, but also close to the metal-insulator transition has been discussed by different theoretical and numerical approaches in the 
past~\cite{sahebsara2008,yamada2014,laubach2015,yoshioka2009,yang2010,antipov2011,shirakawa2017}. The term {\it weak-Mott insulator} has 
been used in this case, namely when a spin liquid intrudes between the weak-coupling metal and the strong-coupling antiferromagnetic 
insulator~\cite{grover2010}. In particular, recent extensive DMRG calculations~\cite{szasz2020,chen2021} highlighted the possibility for
a gapped chiral spin liquid close to the Mott transition. A possible description of such a state has been proposed within a bosonic RVB 
description~\cite{zhang2021}, as well as within a spin model with the four-spin ring-exchange term~\cite{cookmeyer2013}. Calculations 
of Ref.~\cite{chen2021} are limited to 4-leg cylinders, which highly frustrate the 120$^{\circ}$ magnetic pattern, since the corresponding 
${\bf k}$ vectors are not allowed by the quantization of momenta. Instead, in Ref.~\cite{szasz2020} also 6-leg cylinders have been considered, 
even if the presence of two almost degenerate momentum sectors at intermediate $U/t$ can make the interpretation of the results not completely 
trivial. In addition, a recent study at finite temperature, still focusing on 4-leg cylinders, highlighted the concomitant presence of chiral 
correlations and nematic order at finite, but low, temperature and intermediate coupling~\cite{wietek2021}. Instead, a DMRG investigation on 
3-leg cylinders suggested that a gapless and nonchiral spin liquid appears close to the Mott transition~\cite{peng2021}. A more conventional 
picture, with a direct transition between a metal and an insulator with magnetic order, has been found in 
Refs.~\cite{lee2008,watanabe2008,tocchio2013}. We would like to remark that the analysis of the insulating phase in the vicinity of the Mott 
transition is complicated by the significant difference in locating the Mott transition, as observed with the different methods. Finally, the 
effect of NNN hopping has been addressed in Ref.~\cite{misumi2017}, using the VCA method with few ($12$) sites, leading to a large spin-liquid 
region for $t^\prime/t>0$, and in a VMC study that suggests always a direct transition between a metal and a magnetic insulator, even if with 
an asymmetry between positive and negative values of $t^\prime/t$~\cite{tocchio2020}.

In this work, we present variational Monte Carlo results, based upon Jastrow-Slater wave functions and backflow correlations, for the Hubbard 
model on the triangular lattice on $N$-leg cylinders, with $N=4$, $5$, and $6$. The role of the NNN hopping term $t^\prime$ is also discussed. 
On 4-leg cylinders, we find a chiral spin liquid on a relatively extended region in the vicinity of the metal-insulator transition, when 
$t^\prime/t<0$. This intermediate phase is gapped, in analogy with DMRG results. In addition, a gapless nonchiral state appears at larger values 
of the Coulomb repulsion. For $t^\prime/t \ge 0$, the Mott insulator is always nonchiral, with quasi-long-range 120$^\circ$ magnetic order, even 
if the chiral spin-liquid state is quite close in energy, at least in the vicinity of the Mott transition. On 5- and 6-leg cylinders, the chiral
spin liquid does not give the best variational energy. For $N=5$, the whole insulating regime is described by a gapless nonchiral spin liquid, 
while for $N=6$ the phase diagram is similar to the one found in the two-dimensional case~\cite{tocchio2020}, with antiferromagnetic order close 
to the metal-insulator transition and a gapless nonchiral spin-liquid phase at strong coupling.

The paper is organized as follows: in section~\ref{sec:method}, we describe the model and the various variational wave functions, as well as
the quantities that have been used to obtain the important information; in section~\ref{sec:results}, we present the numerical results; 
finally, in section~\ref{sec:concl}, we draw our conclusions.

\begin{figure}
\includegraphics[width=0.9\textwidth]{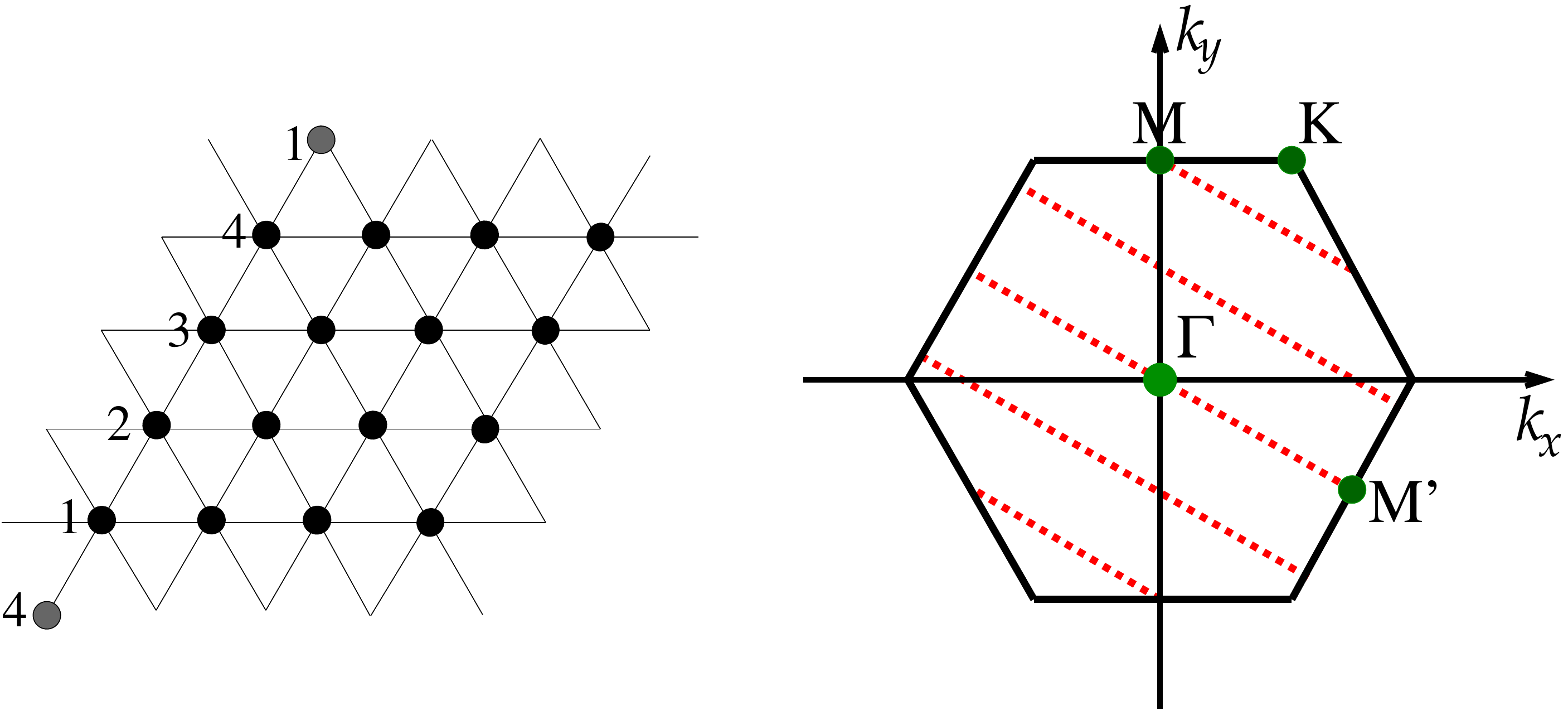}
\caption{Left panel: Triangular lattice on a cylinder with $L_2=4$, with periodic boundary conditions along ${\bf a}_1$ and ${\bf a}_2$. 
This geometry is different from the YC4 one of Refs.~\cite{szasz2020,chen2021}. Right panel: First Brillouin zone of the triangular lattice. 
The red dashed lines indicate the allowed momenta on a cylinder with $L_2=4$.}
\label{fig:lattice}
\end{figure}

\section{Model and method}\label{sec:method}

We consider the single-band Hubbard model on the triangular lattice:
\begin{equation}\label{eq:hubbard}
\begin{split}
{\cal H} = & -t \sum_{\langle i,j\rangle,\sigma} c^\dagger_{i,\sigma} c_{j,\sigma}^{\phantom{\dagger}} 
 - t^\prime \sum_{\langle\langle i,j\rangle\rangle,\sigma} c^\dagger_{i,\sigma} c_{j,\sigma}^{\phantom{\dagger}} + \textrm{H.c.} \\
           & +U \sum_{i} n_{i,\uparrow} n_{i,\downarrow}\,,
\end{split}
\end{equation}
where $c^\dagger_{i,\sigma}$ ($c^\dagga_{i,\sigma}$) creates (destroys) an electron with spin $\sigma$ on site $i$ and 
$n_{i,\sigma}=c^\dagger_{i,\sigma} c^\dagga_{i,\sigma}$ is the electronic density per spin $\sigma$ on site $i$. The NN and NNN hoppings 
are denoted as $t$ and $t^\prime$, respectively; $U$ is the on-site Coulomb interaction. We define three vectors connecting NN sites, 
${\bf a}_1=(1,0)$, ${\bf a}_2=(1/2,\sqrt{3}/2)$, and ${\bf a}_3=(-1/2,\sqrt{3}/2)$; in addition, we also define three vectors for NNN sites, 
${\bf b}_1={\bf a}_1+{\bf a}_2$, ${\bf b}_2={\bf a}_2+{\bf a}_3$, and ${\bf b}_3={\bf a}_3-{\bf a}_1$. We consider clusters with periodic 
boundary conditions defined by ${\bf T}_1=L_1 {\bf a}_1$ and ${\bf T}_2=L_2 {\bf a}_2$, in order to have $L=L_1 \times L_2$ sites. We focus 
on cylinders with four ($L_2=4$), five ($L_2=5$), and six ($L_2=6$) legs, see the case with $L_2=4$ in Fig.~\ref{fig:lattice}. Most of the 
calculations have been done with $L_1=30$, which is large enough not to suffer from significant finite-size effects. The half-filled case, 
where the Mott transition takes place, is considered. In this case, only the sign of the ratio $t^\prime/t$ is relevant and not the individual 
signs of $t$ and $t^\prime$.

Our numerical results are obtained by means of the VMC method, which is based on the definition of suitable wave functions to approximate the
ground-state properties beyond perturbative approaches~\cite{becca2017}. In particular, we consider the so-called Jastrow-Slater wave functions
that include long-range electron-electron correlations via the Jastrow factor~\cite{capello2005,capello2006}, on top of an uncorrelated Slater
determinant (possibly including electron pairing). In addition, the so-called backflow correlations will be applied to the Slater determinant,
in order to sizably improve the quality of the variational state~\cite{tocchio2008,tocchio2011}. Thanks to Jastrow and backflow terms, these
wave functions can reach a very high degree of accuracy in Hubbard-like models, for different regimes of parameters, including frustrated 
cases~\cite{leblanc2015}. Therefore, they represent a valid tool to investigate strongly-correlated systems, competing with state-of-the-art 
numerical methods, as DMRG or tensor networks.

Our variational wave function for describing the spin-liquid phase is defined as:
\begin{equation}\label{eq:wf_SL}
|\Psi_{\textrm{SL}}\rangle={\cal J}_d |\Phi_{\textrm{MF}}\rangle\,,
\end{equation}
where ${\cal J}_d$ is the density-density Jastrow factor and $|\Phi_{\textrm{MF}}\rangle$ is a state where the orbitals of an auxiliary 
Hamiltonian are redefined on the basis of the many-body electronic configuration, incorporating virtual hopping processes, via the backflow 
correlations~\cite{tocchio2008,tocchio2011}. 

The density-density Jastrow factor is given by
\begin{equation}\label{eq:jastrowd}
{\cal J}_d = \exp \left ( -\frac{1}{2} \sum_{i,j} v_{i,j} n_{i} n_{j} \right )\,,
\end{equation}
where $n_{i}= \sum_{\sigma} n_{i,\sigma}$ is the electron density on site $i$ and $v_{i,j}$ are pseudopotentials that are optimized for every 
independent distance $|{\bf R}_i-{\bf R}_j|$. The density-density Jastrow factor allows us to describe a nonmagnetic Mott insulator for a 
sufficiently singular Jastrow factor $v_q \sim 1/q^2$ ($v_q$ being the Fourier transform of $v_{i,j}$)~\cite{capello2005,capello2006}.

The auxiliary Hamiltonian is then defined as follows:
\begin{equation}\label{eq:H_BCS}
{\cal H}_{\textrm{MF}} = \sum_{k,\sigma} \xi_k c^{\dagger}_{k,\sigma}c^{\dagga}_{k,\sigma}
+\sum_{k}\Delta_k c^{\dagger}_{k,\uparrow}c^{\dagger}_{-k,\downarrow}+\textrm{H.c.}\,,
\end{equation}
where $\xi_k=\tilde{\epsilon}_k-\mu$ defines the free-band dispersion (including the chemical potential $\mu$) and $\Delta_k$ is the singlet 
pairing amplitude. In our previous work on the two-dimensional lattice~\cite{tocchio2020}, we found that the best spin liquid has a nematic 
character, the hopping terms being given by:
\begin{equation}\label{eq:hopp}
\begin{split}
\tilde{\epsilon}_k = & -2t \left[ \cos({\bf k} \cdot {\bf a}_1) + \cos({\bf k} \cdot {\bf a}_3) \right] -2\tilde{t_d} \cos({\bf k}\cdot {\bf a}_2) \\
& -2 \tilde{t}^\prime \left[ \cos({\bf k} \cdot {\bf b}_1) + \cos({\bf k} \cdot {\bf b}_2) + \cos({\bf k} \cdot {\bf b}_3) \right ]\,.
\end{split}
\end{equation}
Instead, the pairing amplitudes are:
\begin{equation}\label{eq:pair}
\Delta_k=2\Delta \left[ \cos({\bf k} \cdot {\bf a}_1) -\cos({\bf k} \cdot {\bf a}_3) \right]+2\Delta_d(\cos{\bf k}\cdot {\bf a}_2)\,,
\end{equation}
which possess a $d$-wave symmetry on the two bonds with hopping $t$. In two dimensions, we found $\tilde{t}_d \approx 0$ and $\Delta_d \approx 0$, 
while on cylinders they may assume finite values. Remarkably, this choice (with different couplings along ${\bf a}_2$ and ${\bf a}_3$) gives
the best variational energy also on cylinders, this implying an explicit breaking in point-group symmetries. 

In addition, we focus on chiral spin-liquid states, which have been claimed to be relevant both in the Heisenberg limit~\cite{song2020} and 
in the Hubbard model close to the Mott transition~\cite{szasz2020,chen2021}. The CSL2 state is a projected $d+id$ superconductor characterized 
by uniform (real) hopping along NN and NNN bonds and a pairing 
\begin{equation}\label{eq:chiral}
\Delta_k=2\Delta \left[ \cos({\bf k} \cdot {\bf a}_1)+\omega \cos({\bf k} \cdot {\bf a}_2) + 
\omega^2 \cos({\bf k} \cdot {\bf a}_3)\right]\,,
\end{equation}
where $\omega=e^{2i\pi/3}$. Another chiral state (dubbed here CSL3) may be defined by the hopping amplitude of Eq.~(\ref{eq:hopp}) and a 
different $d+id$ pairing structure:
\begin{equation}\label{eq:chiral_broken}
\Delta_k=2\Delta \left[ \cos({\bf k} \cdot {\bf a}_1) -\cos({\bf k} \cdot {\bf a}_3) \right]+2i\Delta_d \cos({\bf k} \cdot {\bf a}_2)\,.
\end{equation}
Finally, a chiral spin liquid with $U(1)$ symmetry has been proposed (dubbed CSL1 in Ref.~\cite{song2020}), with magnetic fluxes piercing the 
elementary plaquettes. In presence of density fluctuations, this state breaks the translational symmetry and does not give a competitive 
variational energy. Therefore, in the following, this {\it Ansatz} is not reported.

Within the two-dimensional case, antiferromagnetically ordered wave functions represent an important class of states, since a large portion of 
the phase diagram corresponds to phases that spontaneously break the SU(2) spin symmetry. Cylinders are quasi-one dimensional systems, in 
which a continuous symmetry cannot be broken. Nevertheless, variational wave function can be still constructed from a magnetically ordered 
Slater determinant. Then, density and spin correlations may be inserted by Jastrow factors:
\begin{equation}\label{eq:wf_AF}
|\Psi_{\textrm{AF}}\rangle = {\cal J}_s {\cal J}_d |\Phi_{\textrm{AF}}\rangle\,;
\end{equation}
here, ${\cal J}_d$ is the density-density term of Eq.~(\ref{eq:jastrowd}) and ${\cal J}_s$ is the spin-spin Jastrow factor, which is written 
in terms of a pseudopotential $u_{i,j}$ that couples the $z$-component of the spin operators on different sites
\begin{equation}\label{eq:jastrows}
{\cal J}_s = \exp \left ( -\frac{1}{2} \sum_{i,j} u_{i,j} S^z_{i} S^z_{j} \right )\,.
\end{equation}
Finally, $|\Phi_{\textrm{AF}}\rangle$ is obtained, after taking into account the backflow corrections, from the following auxiliary Hamiltonian:
\begin{equation}\label{eq:AF}
{\cal H}_{\textrm{AF}} = \sum_{k,\sigma}\epsilon_k c^{\dagger}_{k,\sigma}c^{\dagga}_{k,\sigma} +
                         \Delta_{\textrm{AF}} \sum_{i} {\bf M}_i \cdot {\bf S}_i\,,
\end{equation}
where, $\epsilon_k$ is the free dispersion of Eq.~(\ref{eq:hubbard}), ${\bf S}_i=(S^x_i,S^y_i,S^z_i)$ is the spin operator at site $i$ and 
${\bf M}_i$ is defined as ${\bf M}_i=[\cos({\bf Q}\cdot {\bf R}_i),\sin({\bf Q}\cdot {\bf R}_i),0]$, where ${\bf Q}$ is the pitch vector. 
The three-sublattice 120$^{\circ}$ order has ${\bf Q}=(\frac{4\pi}{3},0)$ or $(\frac{2\pi}{3},\frac{2\pi}{\sqrt{3}})$, while the stripe 
collinear order with a two-sublattice periodicity has ${\bf Q}=(0,\frac{2\pi}{\sqrt{3}})$ or ${\bf Q}=(\pi,\frac{\pi}{\sqrt{3}})$. On 6-leg 
cylinders, the pitch vector corresponding to the 120$^{\circ}$ order is allowed by the quantization of momenta; instead, on 4- and 5-leg cases 
it is not allowed and we take the closest possible momentum. On 5 legs, also the pitch vector of the stripe collinear order is not allowed.

In general, the effect of the spin-spin Jastrow factor ${\cal J}_s$ is to reduce the value of magnetic order of the uncorrelated Slater 
determinant~\cite{manousakis1991,becca2000}. In purely one-dimensional systems, the presence of a long-range Jastrow factor is able to 
completely destroy magnetic order, leading to the correct behavior of the spin-spin correlations~\cite{franjic1997}. On cylinders with a 
finite number of legs $N$, a residual magnetic order persists, thus giving rise to a spurious wave function that breaks the SU(2) symmetry. 
Here, we interpret the possibility to stabilize this kind of variational state as the tendency to develop magnetic order in the two-dimensional 
system. For simplicity, in the following, the {\it Ansatz} of Eq.~(\ref{eq:wf_AF}) will be denoted by ``antiferromagnetic''. We remark that, 
in principle, it would be possible to restore the SU(2) symmetry by projecting on the $S=0$ subspace~\cite{mizusaki2004}. However, this 
procedure is rather computationally expensive, whenever the computational basis has a definite value of $S^z=\sum_i S^z_i$ but not of 
$S^2=(\sum_i {\bf S}_i)^2$.

All the pseudopotentials in the Jastrow factors, the parameters in the auxiliary Hamiltonian, as well as the backflow corrections are optimized 
with the stochastic reconfiguration method~\cite{becca2017}.

In order to assess the metallic or insulating nature of the ground state we can compute the static density-density structure factor:
\begin{equation}\label{eq:Nq}
N({\bf q})=\frac{1}{L}\sum_{i,j}\langle n_in_j\rangle^{i {\bf q}\cdot({\bf R}_i-{\bf R}_j)}\,,
\end{equation}
where $\langle\dots\rangle$ indicates the expectation value over the variational wave function. Indeed, density excitations are gapless 
when $N({\bf q})\propto |{\bf q}|$ for $|{\bf q}| \to 0$, while a gap is present whenever $ N({\bf q})\propto |{\bf q}|^2$ for 
$|{\bf q}|\to 0$~\cite{feynman1954,tocchio2011}. Analogously, the presence of a spin gap can be checked by looking at the small-$q$ 
behavior of the static spin-spin correlations~\cite{tocchio2019}:
\begin{equation}\label{eq:Sq}
S({\bf q})=\frac{1}{L}\sum_{i,j}\langle S^z_i S^z_j\rangle^{i {\bf q}\cdot({\bf R}_i-{\bf R}_j)}\,.
\end{equation}

\begin{figure}
\includegraphics[width=0.9\textwidth]{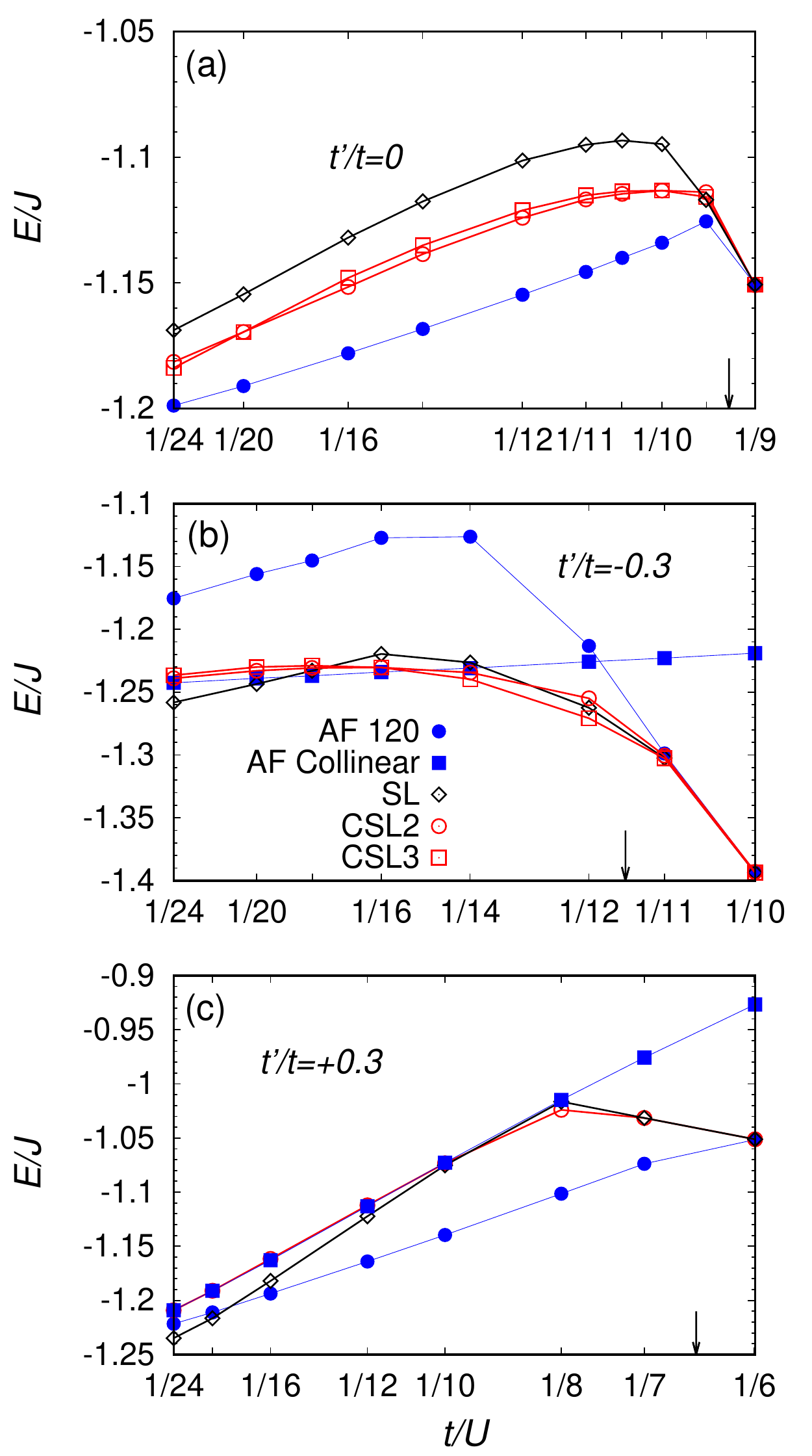}
\caption{Energy (per site) in units of $J=4t^2/U$, as a function of $t/U$ on the 4-leg cylinder for $t^\prime/t=0$ (upper panel), $t^\prime/t=-0.3$ 
(middle panel) and $t^\prime/t=+0.3$ (lower panel). Data are shown for different trial wave functions: The antiferromagnetic state with 
${\bf Q}=(\frac{2\pi}{3},\frac{7\pi}{3\sqrt{3}})$ (blue full circles), the one with ${\bf Q}=(0,\frac{2\pi}{\sqrt{3}})$ (blue full squares), 
the spin liquid (SL) with hopping in Eq.~(\ref{eq:hopp}) and pairing in Eq.~(\ref{eq:pair}) (black empty diamonds), the CSL2 with uniform hopping 
and pairing given by Eq.~(\ref{eq:chiral}) (red empty circles), and the CSL3 defined by Eqs.~(\ref{eq:hopp}) and~(\ref{eq:chiral_broken}) (red 
empty squares). Black arrows denote the metal-insulator transitions. Data are shown for a $L=30 \times 4$ lattice size. Error bars are smaller 
than the symbol size.}
\label{fig:energy_4legs}
\end{figure}

\begin{figure}
\includegraphics[width=0.9\textwidth]{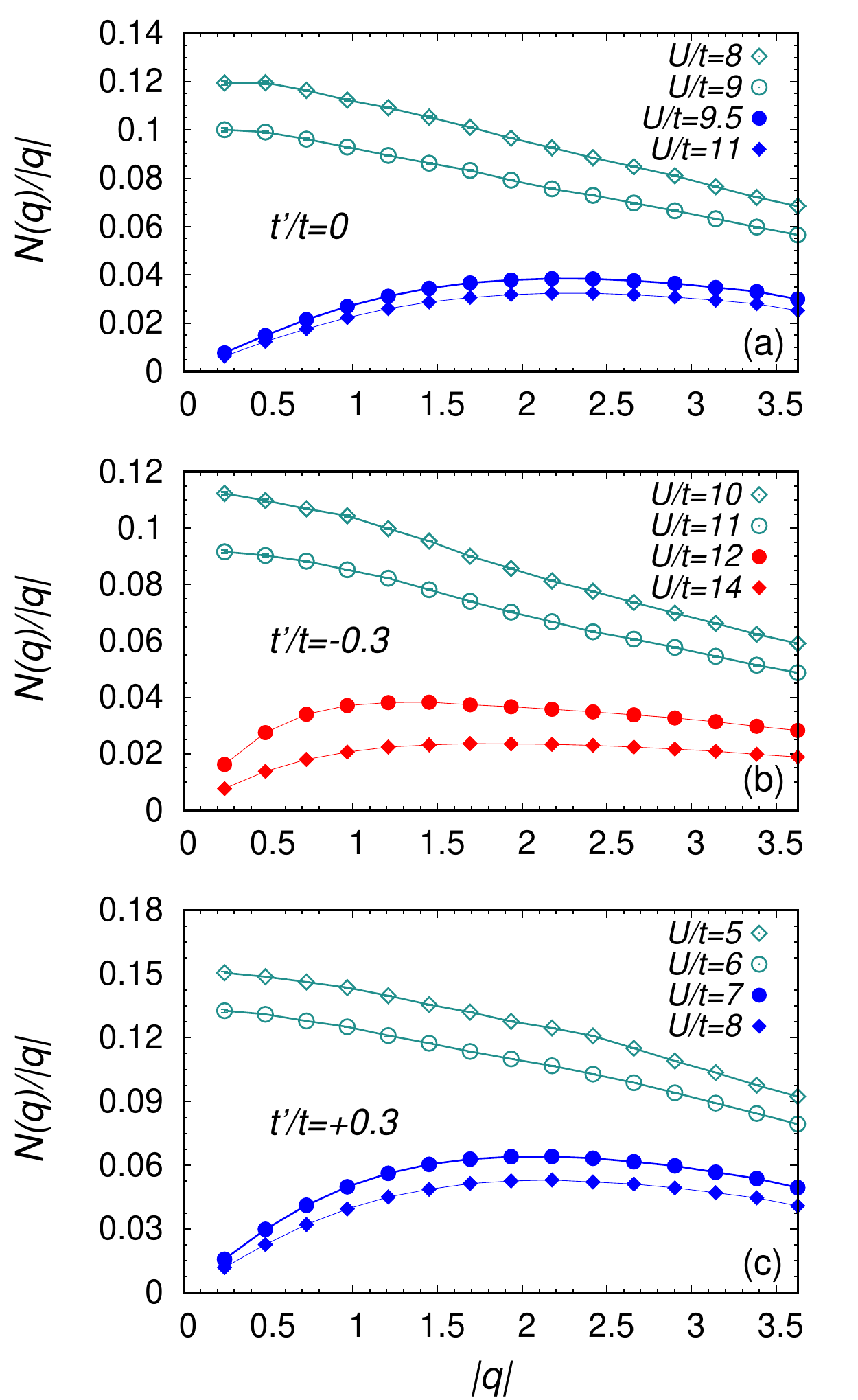}
\caption{Static density-density structure factor $N({\bf q})$, divided by $|{\bf q}|$, computed for the lowest-energy states at different values 
of $U/t$. Results for $t^\prime/t=0$ (upper panel), $t^\prime/t=-0.3$ (middle panel) and $t^\prime/t=+0.3$ (lower panel) are shown on a 4-leg 
cylinder with $L=30 \times 4$, along the line connecting $\Gamma=(0,0)$ to $M'=(\pi,-\frac{\pi}{\sqrt{3}})$. Error bars are smaller than the 
symbol size.}
\label{fig:density}
\end{figure}

\section{Results}\label{sec:results}

Here, we discuss the results for the variational energy of different states on the 4-leg cylinder geometry. Let us start from the case with 
$t^\prime/t=0$, see Fig.~\ref{fig:energy_4legs} (upper panel). In this case the Mott transition occurs between $U/t=9$ and $9.5$, as extracted 
from the low-$q$ behavior of the density-density correlations, see Fig~\ref{fig:density} (upper panel). In the small-$U$ regime, the pairing 
terms $\Delta_k$ [in the spin-liquid {\it Ans\"atze} of Eq.~(\ref{eq:wf_SL})] or the antiferromagnetic parameter $\Delta_{\textrm{AF}}$ [in 
the magnetic wave function of Eq.~(\ref{eq:wf_AF})] are very small and there is a tiny difference among all different variational states. 
This fact indicates that the conducting phase is a standard metal, with neither magnetic nor superconducting order. Instead, in the insulating 
phase, the optimal wave function is the antiferromagnetic one with the pitch vector corresponding to approximately 120$^\circ$ order, i.e., 
${\bf Q}=(\frac{2\pi}{3},\frac{7\pi}{3\sqrt{3}})$. The overall situation is not much different from what has been obtained, within the same 
approach, in the two-dimensional limit~\cite{tocchio2020} (except the fact that in the latter case, a true antiferromagnetic order settles down). 
We also remark that the energy gain of the antiferromagnetic state with respect to the spin-liquid one is smaller on four legs than in two 
dimensions. 

Then, a large spin-liquid region appears immediately above the Mott transition, by including a finite NNN hopping $t^\prime/t=-0.3$, see 
Fig.~\ref{fig:energy_4legs} (middle panel). Here, the metal-insulator transition takes place at $U/t=11.5 \pm 0.5$, see Fig~\ref{fig:density}
(middle panel). The best variational state, between $U/t=12$ and $16$, is given by the CSL3, even though the other spin-liquid states are very 
close in energy. By increasing the ratio $U/t$, the ground state passes through an intermediate phase where the best variational state is the 
antiferromagnetic one (with collinear order), before entering a further (strong-coupling) spin-liquid region that has no chiral features, in 
analogy with the results previously obtained in two dimensions~\cite{tocchio2020}. In Fig.~\ref{fig:energy_4legs}, we do not report the flux 
phase CSL1, since its variational energy is always significantly higher than the other states. For example, at $U/t=12$, we have $E/J=-0.8907(3)$ 
for $t^\prime=0$ and $E/J=-0.9294(3)$ for $t^\prime/t=-0.3$. The situation is radically different by taking the opposite sign of the hopping 
amplitudes, i.e., $t^\prime/t=+0.3$, see Fig.~\ref{fig:energy_4legs} (lower panel). The Mott transition lowers down to $U/t=6.5 \pm 0.5$ [see 
Fig.~\ref{fig:density} (lower panel)], with the insulating state being approximated by the antiferromagnetic state, with pitch vector
${\bf Q}=(\frac{2\pi}{3},\frac{7\pi}{3\sqrt{3}})$, up to $U/t \approx 20$. For larger values of the electron-electron repulsion, a nonchiral 
spin-liquid state emerges. Note that the energies reported for the antiferromagnetic state with collinear order below the Mott transition 
correspond to a local minimum with insulating features.

\begin{figure}
\includegraphics[width=0.9\textwidth]{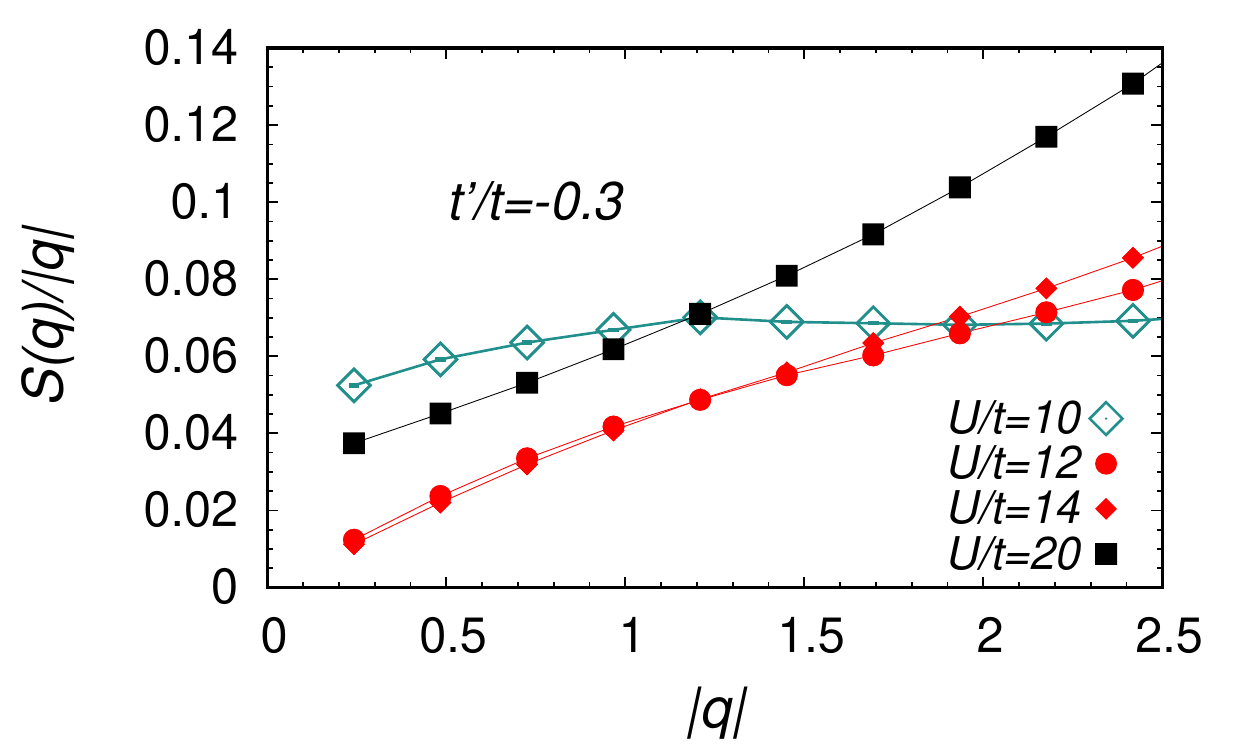}
\caption{Static spin-spin structure factor $S({\bf q})$, divided by $|{\bf q}|$, computed for the lowest-energy states at different values of
$U/t$, for $t^\prime/t=-0.3$. Data are shown on a 4-leg cylinder with $L=30 \times 4$, in the metallic region (green empty diamonds), for the 
chiral spin liquid state CSL3 (red full points) and for the large-$U$ spin liquid (black full squares). The ${\bf q}$ points are located along 
the line connecting $\Gamma=(0,0)$ to $M'=(\pi,-\frac{\pi}{\sqrt{3}})$. Error bars are smaller than the symbol size.}
\label{fig:spin}
\end{figure}

\begin{figure}
\includegraphics[width=0.9\textwidth]{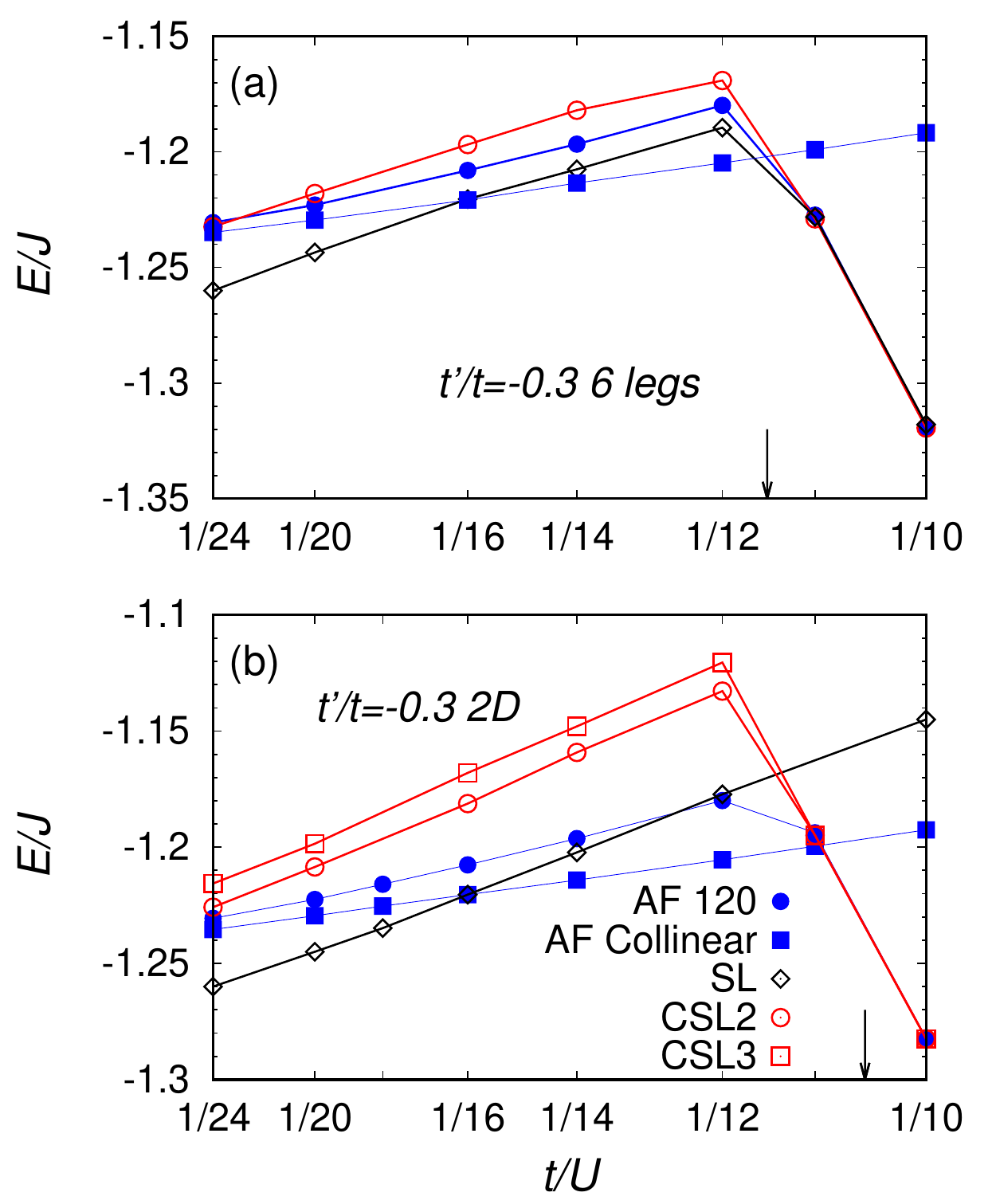}
\caption{Energy (per site) in units of $J=4t^2/U$, as a function of $t/U$, for $t^\prime/t=-0.3$ on the 6-leg cylinder with $L=30 \times 6$ 
(upper panel). The results for the two-dimensional cluster with $L=18 \times 18$ are also reported for comparison (lower panel). Data are shown 
for different trial wave functions: The antiferromagnetic state with 120$^{\circ}$ magnetic order (blue full circles), the state with collinear 
order (blue full squares), the spin liquid with hopping in Eq.~(\ref{eq:hopp}) and pairing in Eq.~(\ref{eq:pair}) (black empty diamonds), the 
CSL2 with uniform hopping and pairing given by Eq.~(\ref{eq:chiral}) (red empty circles), and the CSL3 defined by Eqs.~(\ref{eq:hopp}) 
and~(\ref{eq:chiral_broken}) (red empty squares). Black arrows denote the metal-insulator transitions. Error bars are smaller than the symbol 
size.}
\label{fig:energy_6legs}
\end{figure}

In order to determine the nature of the chiral spin-liquid state, we analyze the spin-spin correlations by computing the spin-spin structure 
factor of Eq.~(\ref{eq:Sq}). In Fig.~\ref{fig:spin}, we report calculations for $t^\prime/t=-0.3$ and values of $U/t$ across the Mott transition. 
The main result is that the chiral spin liquid, realized close to the Mott transition, has a spin gap, since $S({\bf q}) \propto |{\bf q}|^2$
for small values of the momentum ${\bf q}$. This is in agreement with recent DMRG studies~\cite{szasz2020,chen2021}. We remark that this 
feature is solid, since it is also shared by the other two spin-liquid states with nearby energies, i.e., the CSL2 and the nonchiral one
parametrized by Eqs.~(\ref{eq:hopp}) and~(\ref{eq:pair}). On the contrary, the large-$U$ state is gapless. In this regime the optimal parameters 
$\tilde{t}_d \approx 0$ and $\Delta_d \approx 0$ lead to a gapless spectrum in the auxiliary Hamiltonian~(\ref{eq:H_BCS}), thus indicating that 
the nature of the unprojected state is not changed when including the Jastrow factor.

The optimal chiral spin liquid (close to the Mott transition), as well as the nonchiral one (in the strong-coupling regime) are very anisotropic, 
as shown by computing the nearest-neighbor spin-spin correlations $D_j=\langle S^z_{\textbf{R}_i}S^z_{\textbf{R}_i+\textbf{a}_j}\rangle$, with 
$j=1,2,3$. For example, for $t^\prime/t=-0.3$ and $U/t=12$, the CSL3 state has $D_1=D_3=-0.029(1)$ and $D_2=-0.069(1)$. For $U/t=20$, the nonchiral 
spin liquid has $D_1=D_3=-0.101(1)$ and $D_2=+0.041(1)$. Within the error bar, these results are the same from $L=18 \times 4$ to $L=30 \times 4$. 
As discussed in section~\ref{sec:method}, this anisotropy follows directly the parametrization of the spin-liquid state, see Eqs.~(\ref{eq:hopp}) 
and~(\ref{eq:pair}) for the nonchiral {\it Ansatz} and Eqs.~(\ref{eq:hopp}) and~(\ref{eq:chiral_broken}) for the CSL3.  

Then, we show the stability of the chiral spin liquid when going from $N=4$ to $N=6$. Results are shown in Fig.~\ref{fig:energy_6legs}, 
together with the ones for a truly two-dimensional cluster (with $L=18 \times 18$ sites), which has been already discussed in our previous 
work~\cite{tocchio2020}. On a two-dimensional cluster, the CSL3 state is a local minimum, with energy higher than the other states. Instead,
on 6-leg cylinders, the CSL3 state is not reported, since, upon optimization, it converges to the non-chiral state. The most important fact
is that no chiral phases are present in the insulating region close to the metal-insulator transition (which, for $N=6$ appears between $U/t=11$
and $U/t=12$). Here, the insulating phase is either an antiferromagnet with collinear order, in the vicinity of the Mott transition, or a gapless
nonchiral spin liquid, in the strong-coupling regime. Note that, also in this case, the antiferromagnetic state with collinear order becomes a
local minimum below the Mott transition. The reason for the stabilization of the chiral state on 4-leg cylinders comes from its remarkable energy 
gain when going from $N=6$ (or equivalently two dimensions) to $N=4$; by contrast, the variational energies of the antiferromagnetic state do 
not change much when varying $N$. Overall, the resulting phase diagram for $N=6$ is qualitatively similar to the one obtained in two dimensions. 
Therefore, within our approach, the chiral spin liquid exists only for particular values of $N$, like on 4-leg cylinders. 

\begin{figure}
\includegraphics[width=0.9\textwidth]{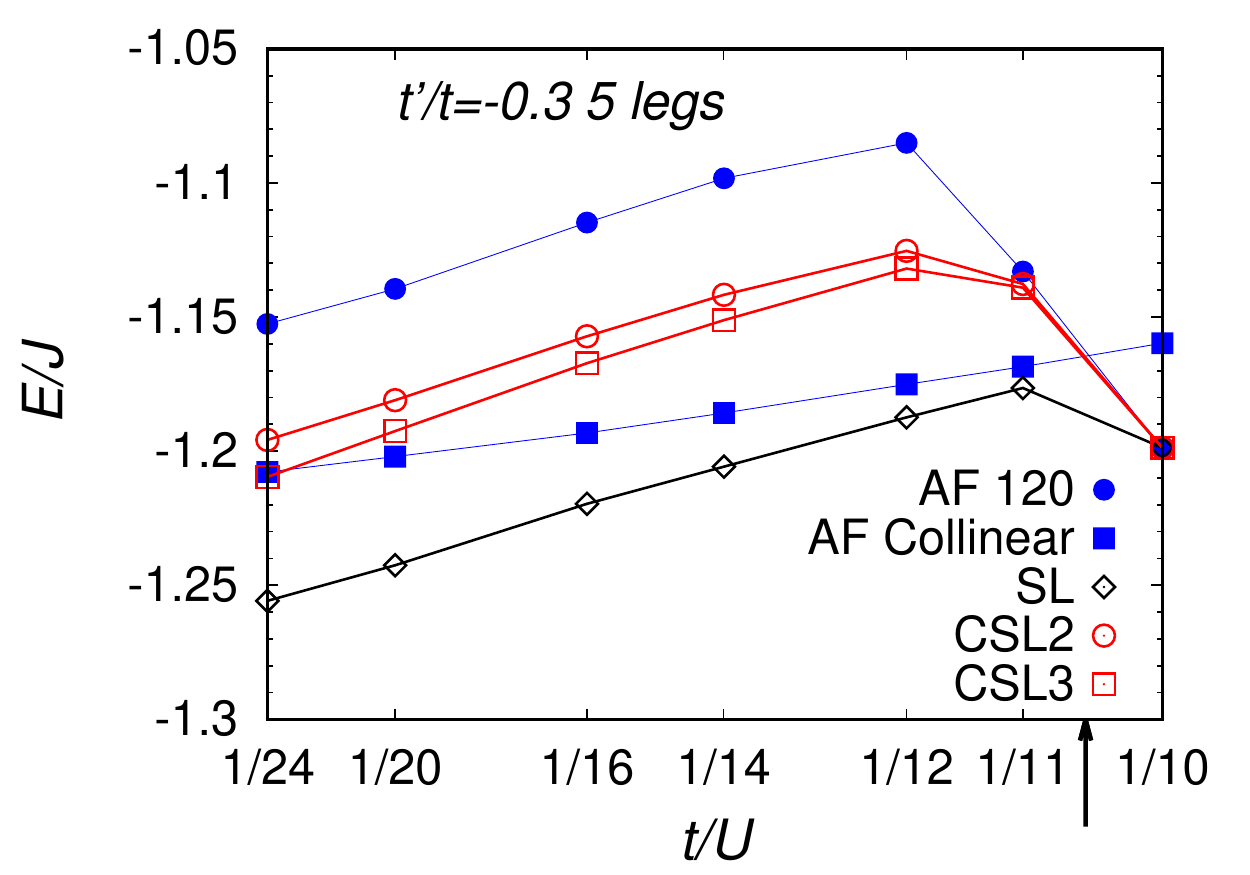}
\caption{Energy (per site) in units of $J=4t^2/U$, as a function of $t/U$, for $t^\prime/t=-0.3$ on the 5-leg cylinder with $L=30 \times 5$. 
Data are shown for different trial wave functions: The antiferromagnetic state with ${\bf Q}=(\frac{2\pi}{3},\frac{26\pi}{15\sqrt{3}})$ (blue 
full circles), the antiferromagnetic state with ${\bf Q}=(\pi,\frac{3\pi}{5\sqrt{3}})$ (blue full squares), the spin liquid with hopping in 
Eq.~(\ref{eq:hopp}) and pairing in Eq.~(\ref{eq:pair}) (black empty diamonds), the CSL2 with uniform hopping and pairing given by 
Eq.~(\ref{eq:chiral}) (red empty circles), and the CSL3 defined by Eqs.~(\ref{eq:hopp}) and~(\ref{eq:chiral_broken}) (red empty squares). 
The black arrow denotes the metal-insulator transition. Error bars are smaller than the symbol size.}
\label{fig:energy_5legs}
\end{figure}

\begin{figure}
\includegraphics[width=0.9\textwidth]{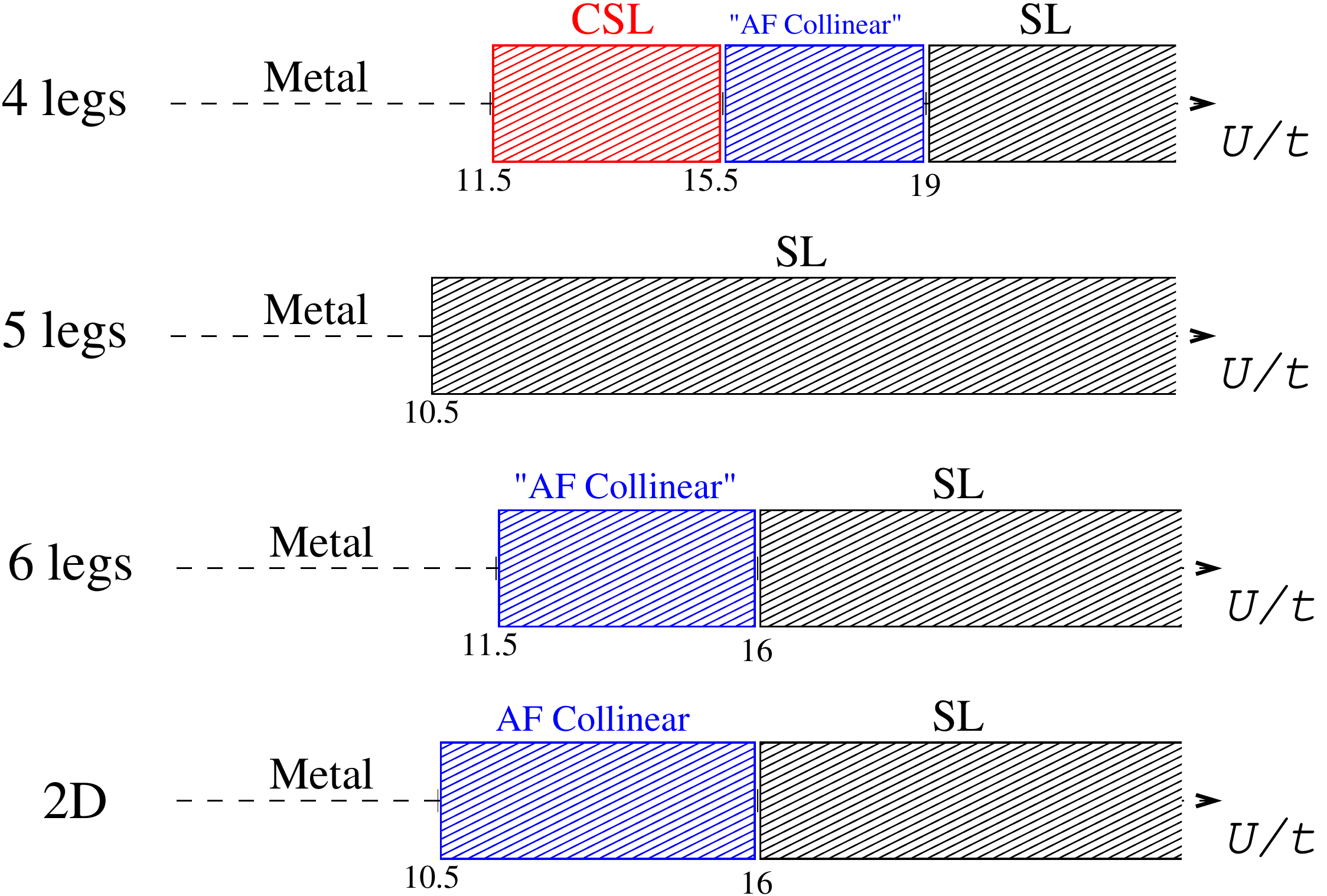}
\caption{Schematic phase diagram of the triangular lattice at $t^\prime/t=-0.3$, as a function of $U/t$. Four different lattices are considered,
with 4, 5, and 6 legs, as well as a two dimensional cluster. The symbol ``SL'' denotes a gapless spin liquid, while the symbol ``CSL'' denotes a 
chiral gapped spin liquid. The antiferromagnetic collinear order on cylindrical geometries is denoted in quotation marks, since no true long-range 
magnetic order can occur.}
\label{fig:pd}
\end{figure}

Finally, we have also considered cylinders with an odd number of legs, i.e., with $L_2=5$. This is a particularly frustrated case, since both 
120$^\circ$ and stripe collinear magnetic correlations are not allowed by the quantization of transverse momenta. Results for the energies of the 
different variational states are reported in Fig.~\ref{fig:energy_5legs}. The Mott transition is determined also in this case by looking at the 
static structure factor of Eq~(\ref{eq:Nq}). In this case, the insulator is always a gapless spin liquid, with no chiral features. Indeed, the 
best variational state is the one defined by Eqs.~(\ref{eq:hopp}) and~(\ref{eq:pair}), with optimal variational parameters $\Delta_d \approx 0$ 
and $\tilde{t}_d \approx 0$, which is the same as the large-$U$ spin liquid reported on the 4-leg case. The two magnetic states are now both 
disfavored because of the 5-leg geometry and they are approximated by the pitch vectors ${\bf Q}=(\pi,\frac{3\pi}{5\sqrt{3}})$ (for the stripe 
collinear order) and ${\bf Q}=(\frac{2\pi}{3},\frac{26\pi}{15\sqrt{3}})$ (for the 120$^\circ$ order). The two chiral states (CSL2 and CSL3) have 
also higher energies with respect to the nonchiral one. Our finding is in agreement with what reported by DMRG in Ref.~\cite{szasz2020}, where 
no chiral features are observed on the 5-leg cylinder when using periodic boundary conditions.

\section{Conclusions}\label{sec:concl} 

In summary, we have studied the Hubbard model on cylinders with a triangular lattice geometry by means of the VMC approach. Both a NN hopping $t$ 
and a NNN hopping $t^\prime$ are considered in the model. First, we focused on the 4-leg case, with different values of the ratio $t^\prime/t$. 
For $t^\prime/t < 0$, a spin liquid is stabilized in the vicinity of the Mott transition. This state is a gapped chiral spin liquid that also
breaks the point-group symmetry. At larger values of $U/t$, a further gapless spin liquid appears. For $t^\prime=0$, the insulating region is 
always antiferromagnetic (with approximately 120$^{\circ}$ order), while, for $t^\prime/t > 0$, we observe a gapless spin liquid in the
strong-coupling regime. However, the chiral spin liquid disappears on cylinders with 5 and 6 legs, as well as in the truly two-dimensional case.
In these cases, a gapless spin liquid survives in the large-$U$ region. These results are summarized in Fig.~\ref{fig:pd}.

Our calculations convey two main messages: On one side, the spin liquid that we obtain on the 4-leg cylinder, close to the Mott transition, is 
chiral and spin gapped, in agreement with recent DMRG calculations~\cite{szasz2020,chen2021}. In addition, the best chiral state breaks the 
reflection symmetry, as also suggested in the finite-temperature tensor-network method of Ref.~\cite{wietek2021}. Nevertheless, within variational 
Monte Carlo, an additional NNN hopping is necessary to stabilize the chiral state. On the other side, our results suggest that a chiral spin 
liquid exists only in particular geometries (e.g., the 4-leg cylinder). Instead, on cylinders with 5 and 6 legs (as well as in two dimensions), 
the chiral spin liquid either is not stable upon optimization or has a variational energy that is quite higher than the optimal state. Finally, 
we observe that chiral flux phases (defined on the $2 \times 1$ unit cell) have a variational energy that is not competitive with other wave 
functions. As every variational calculations, our results suffer from an intrinsic bias, given by the choice of the variational {\it Ansatz}; 
still, the Jastrow-Slater state possess a large flexibility, being able to describe a wide variety of different phases, including quantum spin 
liquids, with or without chiral order. The fact that we do not observe a chiral spin liquid in 5- and 6-leg cylinders and in two dimensional 
clusters, suggests that either this state is not present or it cannot be represented by the {\it Ans\"atze} that have been considered here.

\acknowledgments
We thank F. Ferrari for insightful discussions.

\end{document}